\documentclass[aps,prl,twocolumn,amsmath,amssymb,
reprint,%
]{revtex4-1}

\usepackage{graphicx,color}
\usepackage{epstopdf}
\usepackage{dcolumn}
\usepackage{bm}

\begin{document}
\preprint{}

\title{Free energy cascade in gyrokinetic turbulence}
\author{A.~Ba\~n\'on Navarro}
\affiliation{Universit\'e Libre de Bruxelles, Facult\'e des Sciences,
Physique Statistique et Plasmas CP 231, EURATOM Association, Campus Plaine, 1050 Brussels, Belgium}
\author{P.~Morel}
\affiliation{Universit\'e Libre de Bruxelles, Facult\'e des Sciences,
Physique Statistique et Plasmas CP 231, EURATOM Association, Campus Plaine, 1050 Brussels, Belgium}
\author{M.~Albrecht-Marc}
\affiliation{Universit\'e Libre de Bruxelles, Facult\'e des Sciences,
Physique Statistique et Plasmas CP 231, EURATOM Association, Campus Plaine, 1050 Brussels, Belgium}
\author{F.~Merz}
\affiliation{Max-Planck-Institut f\"ur Plasmaphysik, EURATOM Association, 85748 Garching, Germany}
\author{T.~G\"orler}
\affiliation{Max-Planck-Institut f\"ur Plasmaphysik, EURATOM Association, 85748 Garching, Germany}
\author{F.~Jenko}
\affiliation{Max-Planck-Institut f\"ur Plasmaphysik, EURATOM Association, 85748 Garching, Germany}
\author{D.~Carati} 
\affiliation{Universit\'e Libre de Bruxelles, Facult\'e des Sciences,
Physique Statistique et Plasmas CP 231, EURATOM Association, Campus Plaine, 1050 Brussels, Belgium}
\date{\today} 

\begin{abstract}

In gyrokinetic theory, the quadratic nonlinearity is known to play an important role in the dynamics by
redistributing (in a conservative fashion) the free energy between the various active scales. In the
present study, the free energy transfer is analyzed for the case of ion temperature gradient driven
turbulence. It is shown that it shares many properties with the energy transfer in fluid turbulence.
In particular, one finds a forward (from large to small scales), extremely local, and self-similar
cascade of free energy in the plane perpendicular to the background magnetic field. These findings
shed light on some fundamental properties of plasma turbulence, and encourage the development of
large eddy simulation techniques for gyrokinetics.

\end{abstract}

\maketitle



Fully developed turbulence is fundamentally linked to a conservative transfer of (free) energy in
wavenumber space from drive to dissipation scales~\cite{K41}. While the respective cascade dynamics for
simple fluids (described by the Navier-Stokes equation) has been the subject of countless studies and is
fairly well understood, the situation is quite different for turbulent plasmas, both at large scales
(compared to the gyroradii of the particles) -- described in the context of magnetohydrodynamics --
and, in particular, at small scales -- described by the gyrokinetic equations~\cite{Brizard:2007p478}.
The latter case, in which one deals with a gyrocenter distribution function in three spatial dimensions
as well as two velocity space dimensions (the third velocity space coordinate can be removed analytically
in a low-frequency ordering), shall be the focus of the present work.

In three-dimensional Navier-Stokes turbulence, the kinetic energy is conserved by the convective nonlinearity.
It is usually assumed to be injected into the system at the largest scales through mechanical forcing, and to
be dissipated at the smallest scales by viscous effects. The role of the nonlinearity is then to transfer the
kinetic energy from the large scales to the small ones in what is usually referred to as a cascade process. In
the gyrokinetic formalism, on the other hand, the {\em free energy} acts as the quadratic conserved quantity
(see, e.g., Ref.~\cite{0741-3335-50-12-124024} and various references therein).
It is usually injected into the system at large scales via the background density and temperature gradients,
and expected to be dissipated at small (space and/or velocity space) scales. It is anticipated that one role
of the nonlinear term in gyrokinetic turbulence is to transfer the free energy from the largest perpendicular
scales to the smallest ones~\cite{howes,tatsuno,2010arXiv1003.3933T}, but a definitive investigation of the
free energy transfer dynamics in a self-driven, three-dimensional system (which is the standard case for
magnetically confined plasmas) is still lacking and shall be provided for the first time in the present Letter.


Our study is based on numerical solutions of the nonlinear gyrokinetic equations obtained by means of the
{\sc Gene} code~\cite{jenko:1904,dannert:072309,Merz}. Although {\sc Gene} is able to treat an arbitrary
number of fully gyrokinetic particle species as well as general toroidal geometry, magnetic field fluctuations,
and collisions, these features shall not be used here. Instead, we will focus on the reduced problem of a single
ion species, adiabatic electrons, electrostatic fluctuations, and a large aspect-ratio, circular cross-section
model equilibrium. For this simplified case, the respective (appropriately normalized) equations read (for
details, see Ref.~\cite{Merz}):
\begin{align}
&\frac{\partial f_{j}}{\partial t}  +\Big[ \omega_{n_j} + \left( v_{\parallel}^{2} + \mu B_{0} - \frac{3}{2}\right)\omega_{T_j}\Big]F_{0j}\frac{\partial \bar{\phi_{1}}}{\partial y} \nonumber \\   
&+ \frac{T_{0j}( 2 v_{\parallel}^{2}  + \mu B_{0})}{q_{j}B_{0}}\left(\mathcal{K}_{x} \frac{\partial h_{j}}{\partial x} + \mathcal{K}_{y}\frac{\partial h_{j}}{\partial y} \right) \nonumber \\
&+\left[\bar{\phi_{1}},h_{j}\right]_{x y} + \frac{v_{Tj}}{2}\left[ v_{\parallel}^{2} + \mu B_{0},h_{j}\right]_{z v_{\parallel}} = 0 \,.
\label{GK}
\end{align}
Here, the total distribution function $F_{j}$ of species $j$ is split into a Maxwellian part
$F_{0j} = \pi^{-3/2}e^{-({v_{\parallel}}^{2}+\mu B_{0})}$ and a perturbed part $f_{j}$,
and the nonadiabatic part of $f_{j}$ is given by $h_{j}=f_{j} + (q_{j}\bar{\phi_{1}}/T_{0j})\,F_{0j}$
where $\overline{{\phi}}_1$ is the gyro-averaged electrostatic potential. $h_j$ and
$f_j$ depend on the gyrocenter position ${\bf r}=(x,y,z)$, the parallel velocity $v_\parallel$, the magnetic
moment $\mu$, and the time $t$. As indicated already above, all simulations in this paper are performed in
$\hat{s}-\alpha$ geometry~\cite{s-a-model} with $\alpha=0$, for which the curvature terms are given by
$\mathcal{K}_{x}=-2\sin z$ and $\mathcal{K}_{y}=-2(\cos z + \hat{s}z \sin z)$. Furthermore,
$v_{Tj}=(2T_{j0}/m_{j})^{1/2}$ is the thermal velocity, $\omega_{n_j}=-R\,\partial\log n_{0j}/\partial x$
and $\omega_{T_j}=-R\,\partial\log T_{0j}/\partial x$ are the normalized background density and temperature
gradients, $m_j$ and $q_j$ are the mass and charge of species $j$. The equilibrium magnetic field is taken
to be $B=B_{0}\,B_{\rm ref}$ where $B_{\rm ref}$ is the reference magnetic field on the magnetic axis. Finally,
the Poisson brackets are defined by
\begin{align}
\left[f,g\right]_{a b} =\frac{\partial f}{\partial a}\frac{\partial g}{\partial b}
- \frac{\partial f}{\partial b} \frac{\partial g}{\partial a}\,.
\end{align}
Note that in Eq.~(\ref{GK}), the second term is responsible for the injection of free energy into the system.
The third through fifth terms are, respectively, the curvature, nonlinear, and parallel terms, none of which
acts as a source or sink of free energy. Since the simulations presented below are done without collision
operator, the numerical scheme used in {\sc Gene} is not dissipative, and a statistical steady state cannot
be reached without some form of dissipation~\cite{krommes:3211}. Here,  hyperdiffusion terms
$\mathcal{D}_{z}$ and $\mathcal{D}_{v_{\parallel}}$ are added to dissipate fine-scale fluctuations in $z$ and
$v_{\parallel}$ (for details, see Ref.~\cite{Pueschel}).

Eq.~(\ref{GK}) is complemented by the gyrokinetic Poisson equation which is used to determine the
self-consistent electrostatic potential:
\begin{align}
\sum_{j}\frac{q_{j}^{2}n_{0j}}{T_{0j}}\left[1 - \Gamma_{0}(b_{j})\right]{\phi}_{1}= \hspace{15truemm}\ \nonumber\\
\ \hspace{15truemm}\sum_j n_{0j} \pi q_j B_0 \int J_0(\lambda_j)f_j\,dv_{\parallel}d\mu\,.
\label{Poisson}
\end{align}
Here, $J_{0}$ is the Bessel function and $\Gamma_{0}(b_{j})=e^{-b_{j}}I_{0}(b_{j})$ with the modified Bessel
function $I_0$. The (dimensionless) arguments $b_{j}$ and $\lambda_{j}$ are defined, respectively, as
\begin{align}
b_j = \frac{v_{T_j}^2}{2\Omega_j^2}k_\perp^2\,,\quad
\lambda_j=\frac {v_{Tj}}{\Omega_{j}}(\mu B_0)^{1/2}k_\perp
\end{align}
where $\Omega_j=(q_jB_0)/(m_jc)$ and $k_\perp$ is the perpendicular wave number.


In the absence of drive and dissipation, the gyrokinetic equations, Eqs.~(\ref{GK}) and (\ref{Poisson}),
are known to conserve the free energy $\cal E$ (see, e.g.,
Refs.~\cite{0741-3335-50-12-124024,2010arXiv1003.3933T}) which
is usually split into two quadratic parts according to ${\cal E}={\cal E}_f+{\cal E}_\phi$ with
\begin{align}
\mathcal{E}_f = \sum_{j}\int d\Lambda\,\frac{T_{0j}}{F_{0j}}\frac{f_j^2}{2}\,,\quad
\mathcal{E}_\phi=\sum_{j}\int d\Lambda\,q_{j}\frac{\bar{\phi_1}f_j}{2}\,.
\label{entropy}
\end{align}
Here, $\int d\Lambda = \int d^3x \int \pi B_0 n_{0j}\,dv_{\parallel} d\mu$ denotes phase-space integration.
The evolution equation for the free energy is given by
\begin{align}
\frac{\partial \mathcal{E}}{\partial t} =
\sum_j\int d{\Lambda}\frac{T_{0j}}{F_{0j}}h_j\frac{\partial f_j}{\partial t} = \mathcal{G} - \mathcal{D}
\label{fec}
\end{align}
in terms of the source term
\begin{align}
\mathcal {G} = & -\sum_{j}\int d{\Lambda}\frac{T_{0j}}{F_{0j}}h_j \nonumber \\
 & \cdot\Big[ \omega_{n} + \left( v_{\parallel}^{2} + \mu B_{0} - \frac{3}{2}\right)\omega_{Tj}\Big]F_{0j}
\frac{\partial \bar{\phi_{1}}}{\partial y}
\label{source}
\end{align}
and the (positive definite) dissipative term
\begin{align}
\mathcal{D}=-\sum_{j}\int d{\Lambda}\frac{T_{0j}}{F_{0j}}h_j
\left(\mathcal{D}_{z}f_{j} + \mathcal{D}_{v_{\parallel}}f_{j}\right)\,.
\label{sink}
\end{align}
The quantity $\cal E$ plays the same role in gyrokinetic turbulence as the kinetic energy in fluid
turbulence~\cite{0741-3335-50-12-124024}.


The transfer of free energy between different modes in the saturated turbulent state is induced by the
nonlinear term. Although it does not affect the global value of the free energy
(numerically, this is satisfied in {\sc Gene} up to machine precision), it can change, e.g., the value
of this quantity associated with particular perpendicular wavenumbers. Following the procedure used for
studying energy transfer in Navier-Stokes and in MHD turbulence~\cite{Debliquy:2005p204,PhysRevE.72.046301,
PhysRevE.72.046302,Carati:2006p213}, we decompose the perpendicular wavevector plane into domains and
measure the free energy transfer between these domains. The set of domains $\{d_\ell\}$ is assumed to be
a partition (no intersection between the domains and all domains together cover the entire plane). The
distribution function and electrostatic potential can then be written as a sum over all contributions
for which the perpendicular wavevectors lie in the domain $d_\ell$. As a consequence of the Parseval
theorem, the free energy can also be split into parts which are associated to the domains $d_\ell$:
\begin{align}
{\cal E}=\sum_\ell {\cal E}^\ell= \sum_\ell {\cal E}_f^\ell+\sum_\ell {\cal E}_\phi^\ell\,.
\end{align} 
In the problem considered hereafter, both the entropy and electrostatic contributions to the free energy
are conserved separately by the nonlinearity ${\cal N}$. It is thus legitimate to consider the entropy
conservation independently from the conservation of the electrostatic energy. The evolution of
${\cal E}_f^\ell$ due to the nonlinear term can be expressed as
\begin{align}
\frac{\partial \mathcal{E}_f^\ell}{\partial t} \Big{|}_{\mathcal{N}} =
\sum_{j} \int  d \Lambda \frac{T_{0j}} {F_{0j}}f_{j}^\ell \frac{\partial f_{j}}{\partial t} \Big{|}_{\mathcal{N}}
\end{align}
where we have used the property $\int d \Lambda f_j^\ell f_j^\ell = \int d \Lambda f_j^\ell f_j$ which is easily
proven and expresses the fact that the contributions $f_j^\ell$ are orthogonal ``vectors'' if their scalar
product is defined as the integration over $\Lambda$. Introducing the explicit form of the nonlinearity,
one obtains
\begin{align}
\frac{\partial \mathcal{E}_f^\ell}{\partial t} \Big{|}_{\mathcal{N}}
= \sum_{j} \int  d \Lambda \frac{T_{0j}} {F_{0j}}f_j^\ell \left[ \bar{\phi}_{1},f_{j} \right]_{x y}
= \sum_{\ell_1,\ell_2} T^{\ell;\ell_1,\ell_2}
\label{doublesum}
\end{align}
where the three-domain interaction terms are defined as
\begin{align}
T_{f}^{\ell;\ell_1,\ell_2}=\sum_{j} \int  d \Lambda\frac{T_{0j}}{F_{0j}} f_j^\ell\left[ \bar{\phi}_{1}^{\ell_1},f_j^{\ell_2}\right]_{xy}\,.
\label{T3f}
\end{align}
Eq.~(\ref{doublesum}) shows that the evolution of the entropy associated to the domain $d_\ell$ is the sum of
triple interactions between wave vectors associated to the domains $d_\ell$, $d_{\ell_1}$ and $d_{\ell_2}$. This
is not a surprise since, like in the Navier-Stokes equation, the  quadratic nonlinearity in the gyrokinetic
equation is responsible for triadic interactions between the Fourier modes. Proposing a clean definition of
the energy transfer between two domains might thus be problematic in such a picture. However, considering the
structure of these three-domain interaction terms, the following two-domain interaction terms is a natural
quantity to investigate:
\begin{align}
T_{f}^{\ell,\ell'}=\sum_{\ell_1}T_{f}^{\ell;\ell_1,\ell'}=\sum_{j} \int  d \Lambda\frac{T_{0j}}{F_{0j}} f_j^\ell\left[ \bar{\phi}_{1},f_j^{\ell'}\right]_{xy}\,.
\label{T2f}
\end{align}
These two-domain interaction terms will be interpreted as the energy transfers between the domains $d_\ell$
and $d_{\ell'}$, even if the redistribution of the free energy between the different domains by the nonlinear
term cannot be fully understood without considering triadic interactions.  As a consequence of the Poisson
bracket structure, it is easy to show that $T_{f}^{\ell,\ell'}=-T_{f}^{\ell',\ell}$, which reinforces the
interpretation in terms of free energy exchange. Indeed, if the domain $d_\ell$ is considered to receive
a certain amount of free energy per unit of time $T_{f}^{\ell,\ell'}$ from the domain $d_{\ell'}$, then the
domain $d_{\ell'}$ is seen as loosing exactly the same amount of free energy per unit of time in profit
of the domain $d_\ell$. The same approach can be used to define three-domain and two-domain interaction
terms for the electrostatic part of the free energy with the following definitions:
\begin{align}
T_{\phi}^{\ell;\ell_1,\ell_2} = \sum_{j} \int d \Lambda \bar{\phi}_1^\ell \left[ \bar{\phi}_1^{\ell_2},f_{j}^{\ell_1} \right]_{x y}\,,\\
T_{\phi}^{\ell,\ell'} =\sum_{\ell_1}T_{\phi}^{\ell;\ell_1,\ell'}=
\sum_{j} \int d \Lambda \bar{\phi}_1^\ell \left[ \bar{\phi}_1^{\ell'},f_{j} \right]_{x y}\,.
\label{Te}
\end{align}
The complete dynamical equation for $\mathcal{E}^\ell$ then reads
\begin{align}
\frac{\partial \mathcal{E}^\ell}{\partial t} = \sum_{\ell'}T_{f}^{\ell,\ell'}+ \sum_{\ell'}T_{\phi}^{\ell,\ell'} +\mathcal{G}^\ell-\mathcal{D}^\ell
\label{completedtEell}
\end{align}
where the source and dissipation terms, $\mathcal{G}^\ell$ and $\mathcal{D}^\ell$, are given, respectively,
by Eqs.~(\ref{source}) and (\ref{sink}), using $h_j^\ell$, $f_j^\ell$, and $\bar{\phi}_{1}^\ell$.


The free energy transfer terms defined above are now evaluated from a numerical simulation using {\sc Gene}.
The physical parameters employed in this context correspond to a widely used standard case of collisionless
ion temperature gradient (ITG) turbulence known as the Cyclone Base Case~\cite{dimits:969}. The simulation
domain is about 125 ion gyroradii wide in the perpendicular directions, and
$256 \times 64 \times 64 \times 32 \times 8$ grid points are used in $(x,y,z,v_\parallel,\mu)$ space.
For further analysis, the perpendicular wavevector plane is divided into shells
$d_\ell=\{ \mathbf{k}_\perp \text{ such as } K_{\ell} <|\mathbf{k}_\perp| \le K_{\ell+1} \}$ where the shell
boundaries $K_\ell$ are chosen to grow algebraically $K_{\ell+1}=\lambda K_\ell$, with $\lambda = 2^{1/5}$
between shell $\ell=3$ and $\ell=24$. The first shell boundaries have been chosen differently
($K_1=0$, $K_2 = 0.2$, $K_3 = 0.3$) in order to ensure that enough modes belong to those shells.
Moreover, in order to limit the number of shells, the last shell ($\ell=25$) is wider and limited by
$K_{25} = 6.3$ and $K_{26}=  | \mathbf{k_{\perp}} |_{\max} = 14.6$.

\begin{figure}[htb!]
\includegraphics{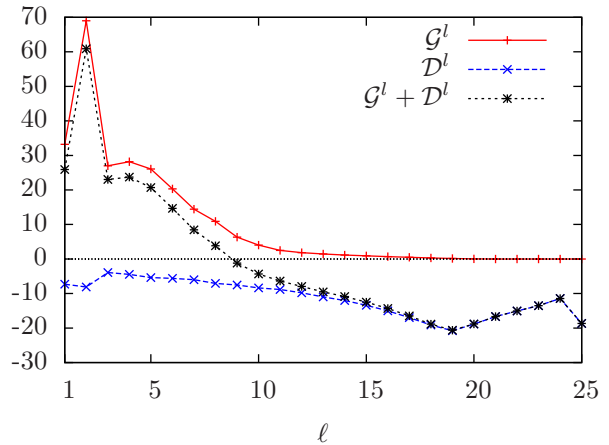}
\caption{Shell decompositions in perpendicular wavenumber space of the drive ($\mathcal{G}^\ell$) and
dissipation ($\mathcal{D}^\ell$) terms (as well as their sum) from a {\sc Gene} simulation of ITG turbulence.}
\label{fig:spectral}
\end{figure}

Fig.~\ref{fig:spectral} shows the numerical results for the source and dissipation terms (averaged over time
during the saturated phase of the simulation). As expected, the injection of free energy is well localized at
low $k_\perp$ . However, as it turns out, the dissipative terms are not just active in the high $k_\perp$  range,
but throughout the entire $k_\perp$  spectrum, including the drive range. An explanation of this phenomenon may
be provided in terms of the nonlinear coupling to damped eigenmodes, as is discussed in Ref.~\cite{Hatch10}.
There is a net source of free energy up to shell $\ell = 9$ and a net dissipation beyond that. The peak in
the dissipation may be due to the fact that the largest shells are not complete because we are
using a discretization in $k_x$ and $k_y$. 

\begin{figure}[h]
\includegraphics{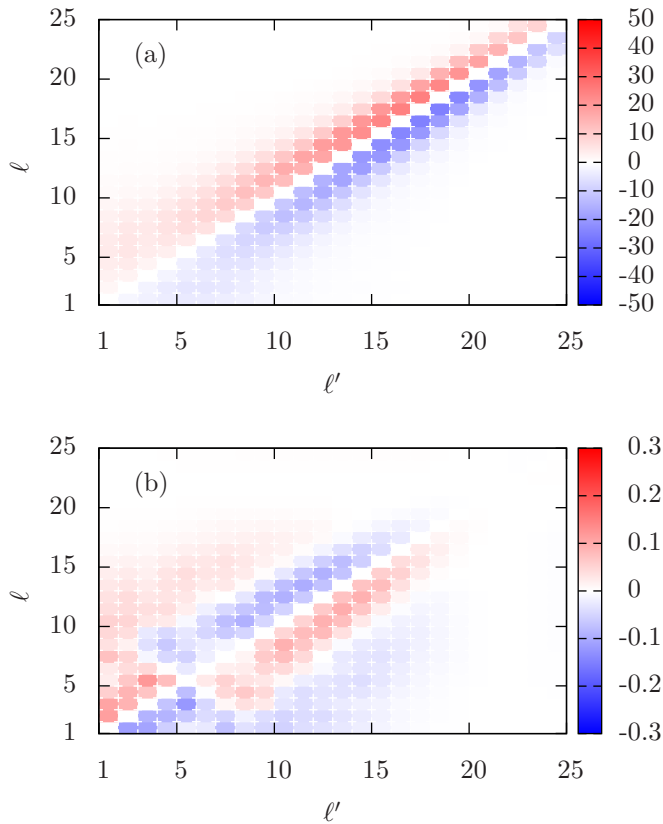}
\caption{Shell-to-shell transfer in perpendicular wavenumber space of entropy (a) and electrostatic energy (b)
from a {\sc Gene} simulation of ITG turbulence.}
\label{fig:transfer}
\end{figure}

The corresponding shell-to-shell free energy transfer terms are shown in Fig~\ref{fig:transfer}, and
various interesting features can be observed there. First, the entropy transfer is larger than the
electrostatic energy transfer by almost two orders of magnitude, thus dominating the total free energy
transfer. Second, the free energy transfer is from the large scales to the small ones (as one might
have expected); the transfer is systematically negative for $\ell'>\ell$ and, due to the antisymmetry
property, systematically positive otherwise. Third, the free energy transfer is very
local in wavenumber space. Indeed, only values of $T_\text{tot}^{\ell,\ell'}$ with $\ell$ close to $\ell'$
are significantly different from zero. In practice, for $|\ell-\ell'|>5$ the free energy transfers
almost vanish. This corresponds to a ratio of wave numbers between the two shells of the order of two.
Fourth, a limited self-similarity range can be identified for $\ell$ between 13 and 20. Indeed, in this
range, the total transfers $T_\text{tot}^{\ell,\ell'}$ seem to depend on $\ell-\ell'$ only, and not on the
two indices separately. Considering the limited resolution of the simulation analysed here, this property
is rather unexpected. Indeed, the analysis of the source and dissipation terms (see Fig.~\ref{fig:spectral})
does not show the existence of a range of scales in which both these terms would be negligible. However, they
are obviously sufficiently small to allow for cascade dynamics to develop (see also Ref.~\cite{Hatch10}).
This is actually to be expected if the nonlinear frequencies characterizing the free energy transfer exceed
the linear ones characterizing the dissipation. 


Interestingly, the spectral transfer of free energy in gyrokinetic turbulence thus exhibits various similarities
with respect to the kinetic energy transfer measured in fully developed Navier-Stokes turbulence, although this
is not all clear {\em a priori}. In particular, there is a (strongly) local, self-similar forward cascade --
despite the absence of an inertial range. Insights like these may be expected
to guide the application of large-eddy simulation techniques~\cite{romo84,agmukn01} to gyrokinetics. Here, the
idea is to only retain the dynamics of the largest scales while the smallest ones are modelled. Indeed, if the
smallest scales are proven to act systematically as a sink of free energy like it was the case here, it is
reasonable to propose a dissipative model for these small scales and consequently reduce as much as possible 
the numerical resolution. On such a basis, it may well become possible to reduce the computational effort for
gyrokinetic turbulence simulations by a significant amount. The present work represents a relevant step in that
direction.

{\em Acknowledgements.}
The authors would like to thank G.~Plunk, T.~Tatsuno and D.~Hatch for very fruitful discussions. This work has been
supported by the contract of association EURATOM - Belgian state. The content of the publication is the
sole responsibility of the authors and it does not necessarily represent the views of the Commission or
its services. D.C. is supported by the Fonds de la Recherche Scientifique (Belgium).  


\end{document}